# Liquefaction-Induced Dam Failure Simulation – A Case for the Material Point Method

Ezra Y. S. Tjung
*Calvin Institute of Technology*

Kenichi Soga
*University of California, Berkeley*

ABSTRACT: Seismic analysis of earthen dams is paramount to evaluate the risk of potential liquefaction and strain softening that can cause flow failure. Even though the current state of the art has moved away from the simplified empirical methods, modeling such large deformation flow failure remains a challenge especially in light of stress/strain history-dependent materials. The Material Point Method (MPM) describes the deformation of a continuum body discretized by a finite number of Lagrangian material points moving through an Eulerian background grid. The MPM is ideal for modeling large deformations with history-dependent constitutive models within the continuum framework. The upstream flow failure of the Lower San Fernando Dam during the 1971 San Fernando Earthquake is used in this paper to demonstrate the advantage of the MPM, where it successfully predicted the final deformed shape.

**Keywords**: large deformation, material point method, dam failure, earthquake, liquefaction

## 1 INTRODUCTION

The failure of the Lower San Fernando Dam during the 1971 San Fernando Earthquake in California has been subject of many interesting studies in the last 50 years (Seed et al., 1973, 1975a, 1975b, 1989; Castro et al., 1985, 1992; Gu et al., 1993; Ming and Li, 2003; Chowdhury et al., 2018; Chowdhury, 2019). The earthen dam experienced liquefaction-induced large deformations with blocky features which resulted in flow failure on its upstream side. Figure 1 is a photo taken after the earthquake event, illustrating the nearly catastrophic upstream slide. Within four hours after the earthquake event, drawdown process of the reservoir was carried out and approximately 80,000 people were immediately evacuated from the area downstream. Figure 2 shows the upstream side of the dam after the draw down. Overall, the sliding mass of the dam travelled approximately 42 m (140 ft) into the reservoir, which is significant considering that the tallest part of the dam was barely 40 m high. This is further illustrated through the cross section presented in Figure 3.

This event emphasized the importance of seismic analysis of earthen dams in evaluating the risk of potential liquefaction and strain softening. The current state of practice has moved away from simplified empirical methods to fully-coupled continuum analyses using the finite difference (FDM) and finite element method (FEM). However, a key limitation of existing mesh-based continuum approaches is their inability to maintain numerical accuracy as dams undergo large deformations, due to mesh-distortion effects. In order to understand the run-out hazard of earthen dams, it is crucial to model the degradation of the soil strength with the deformation especially at large strains.

The Material Point Method (MPM) describes the deformation of a continuum body discretized by a finite number of Lagrangian material points moving through an Eulerian background grid. The MPM is ideal for modeling large deformations of history-dependent materials within the continuum framework (Soga et al., 2016). The hybrid Eulerian-Lagrangian approach employed in the MPM successfully avoids mesh distortion issues (Sulsky et al., 1994, 1995).

This paper presents a preliminary study to illustrate the capability of the MPM in simulating large deformation dam failure within continuum solid mechanics framework with the Lower San Fernando Dam as a case study.

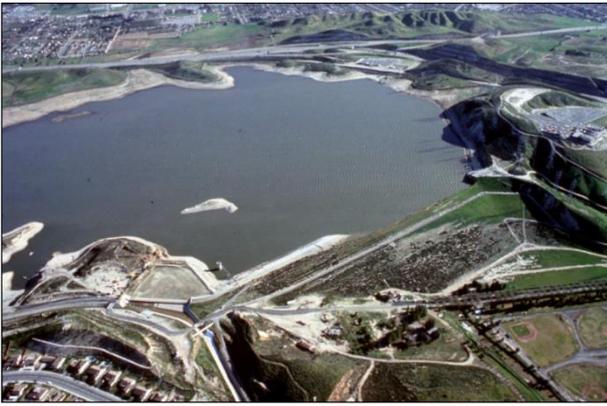

Figure 1. Oblique aerial view of the Lower San Fernando Dam at the bottom of this photo, shortly after the earthquake event and before the resvoir drawdown (Steinburgge Collection, 1971; NICEE Library, University of California Berkeley; after Chowdhury, 2019).

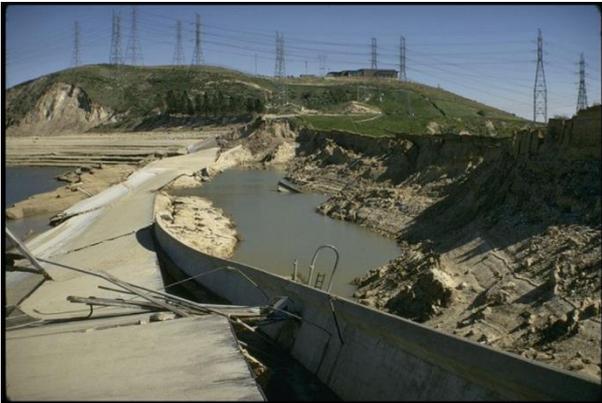

Figure 2. Side view of the Lower San Fernando Dam (Steinburgge Collection, 1971; NICEE Library, University of California Berkeley; after Chowdhury, 2019).

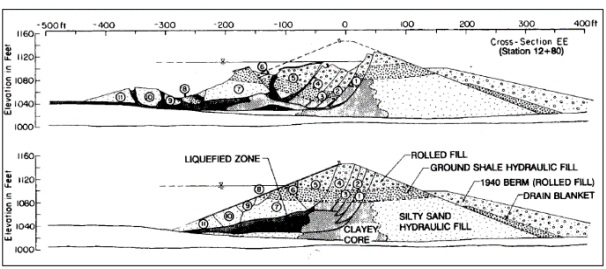

Figure 3. Pre- and post-failure cross-section of the Lower San Fernando Dam after Seed et al., 1973, 1989)

## 2 MATERIAL POINT METHOD

### 2.1 *Theory*

The Material Point Method (MPM) is a hybrid Eulerian-Lagrangian continuum approach for modeling large deformation of history-dependent materials (Sulsky et al., 1994, 1995). Due to its discretization of the domain, the MPM is considered as a particle-based method.

The material domain is discretized into a set of Lagrangian material points, which carry information such as their mass, velocity and other history-dependent material variables. Each material point represents a fixed mass through the MPM computation to ensure mass conservation. Their deformations are determined by the equation of dynamic momentum balance (Newton's law of motions).

The Eulerian background mesh is used purely for computational purposes. Similar to the conventional FEM, the nodes of the mesh are where the discrete governing equations are solved (momentum balance equation). The information required to solve the equations at every time step are mapped from the material points through the shape functions. Once solved, the resulting quantities in the material points are updated from the nodes using the same mapping functions. Note that in the MPM, the information associated with the background mesh is not required and therefore is not stored.

Figure 4 illustrates the computational cycle of the MPM algorithm. This follows the algorithm presented by Sulsky et al. (1995).

### 2.2 *CB-Geo MPM Code Implementation*

CB-Geo is a research group that started with two institutions, namely Cambridge University and the University of California, Berkeley (thus the letters "C" and "B" in the name). CB-Geo focuses on developing open source numerical tools and methods to solve complex geomechanics problems. One of the methods that the research group developins is the MPM.

The objectives of developing the CB-Geo MPM code (Kumar et al., 2019) are: (1) to model large-scale multi-phase multi-physics problems in a reasonable time which requires efficient parallelization and scaling strategy, fitting to be run on High Performance Computing (HPC) environment, (2) to develop a modular system that can handle different MPM algorithms (explicit / implicit / semi-implicit such as projection method), mapping functions (linear / quadratic / GIMP / CPDI / CPDI2) and different constitutive models for different material behavior, and (3) to be able to model complex geometries and boundary conditions.

The study presented in this paper utilizes the CB-Geo MPM code, and it is available for public use through the following hyperlink: https://github.com/cb-geo/mpm.

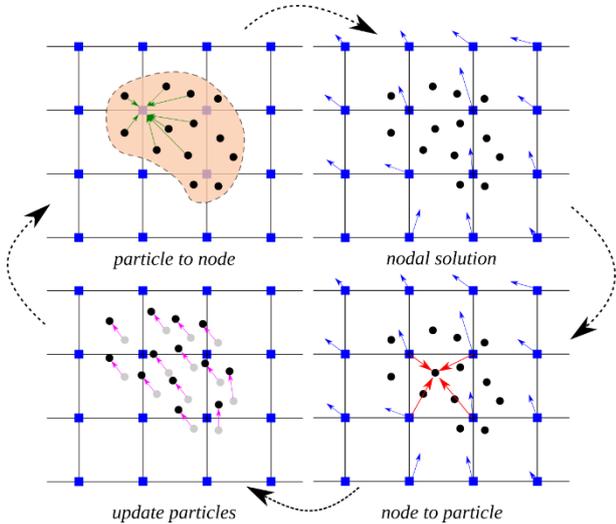

Figure 4. Illustration of the MPM algorithm. (1) A representation of material points overlaid on a background computational mesh. Green arrows represent material point state vectors (mass, volume, velocity, etc.) being projected to the nodes of the computational mesh. (2) The equations of motion are solved onto the nodes, resulting in updated nodal velocities and positions. (3) The updated nodal kinematics is interpolated back to the material points. (4) The location of the material points is updated, and the computational mesh is reset. This figure is after Soga et al. (2016).

## 3 MODELING APPROACH

It is indeed difficult to model the failure of the Lower San Fernando Dam. The case study involves different elements that are worthy to be studied. They include, but not limited to, the ground motion and other seismicity related issues, the behavior of soil under cyclic loading, the modeling of soil under cyclic loading including its stiffness and strength parameters, the triggering of liquefaction, the residual strength of soil post-liquefaction, the strength reduction of soil with large deformation, and the soil-water coupling.

Chowdhury et al. (2018) and Chowdhury (2019) extensively studied the failure of the dam using the commercial software of Itasca's FLAC - Fast Lagrangian Analysis of Continuum, which is based on the Finite Difference Method. Even though they were successful in modeling the inception of the flow failure, their analyses could not usefully predict the final resting geometry despite multiple steps of re-meshing (Figure 5).

The model presented in this paper focuses on the application of the single-phase total stress-based MPM for the post-earthquake failure of the Lower San Fernando Dam under gravity loading, to evaluate the large deformation behavior. The model will not include seismic ground motion or liquefaction triggering criteria, but instead will input residual strength within the liquefied materials as illustrated in Figure 3.

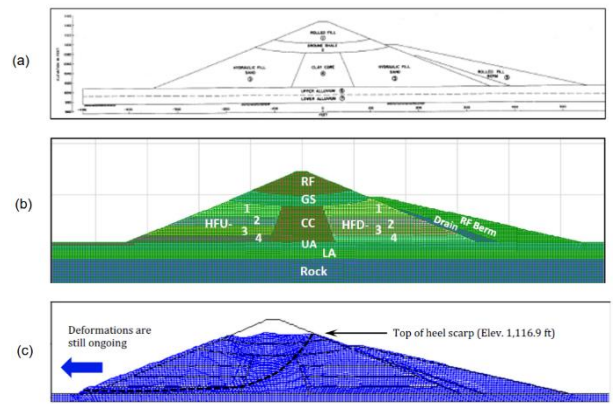

Figure 5. Illustration of the Finite Difference Method (FDM) in modeling a dam failure under seismic load: (a) presents the different materials of the dam, (b) is the FDM mesh, and (c) is the deformed mesh after failure. This study is after Chowdhury (2019).

## 4 LOWER SAN FERNANDO DAM

### 4.1 *MPM Model*

Figure 6 shows the initial configuration of the background mesh and the material points of the MPM model. The Eulerian background mesh consists of 25,000 1-m-by-1-m regular linear quadrilateral elements. Dirichlet boundary conditions are prescribed on the bottom boundary, as well as the left end and the right end of the mesh. In addition, friction coefficient of 0.4 is applied on the bottom of the boundary to model the interface between the bottom of the dam to the foundation materials.

There are 21,823 Lagrangian material points in this simulation. They are equally spaced in both vertical and horizontal directions (0.25 m). Therefore, there are four material points within one element (particles-per-cell).

The constitutive model used in the simulation is the elasto-plastic Mohr-Coulomb model without softening. The material points are divided into eight different materials: (1) rolled fill, (2) ground shale, (3) clay core, (4) upper hydraulic fill, (5) lower hydraulic fill, (6) drain material, (7) berm material, and (8) the liquefied material. For simplicity of the simulation, constant density ($\gamma$) of 1800 kg/m$^3$, Young's modulus (E) of 2 MPa, and Poisson's ratio ($\eta$) of 0.3 are assumed for drained materials. The strength parameters are summarized in Table 1. Though simplified, these values are aligned with the findings of Chowdhury (2019).

In terms of the simulation, the explicit MPM algorithm is used with Update Stress First (USF) scheme. The material point velocity at each step is updated through nodal acceleration. The time step ($\Delta t$) is 0.001 s and the simulation goes for 50 s. A gravity loading of 9.81 m/s$^2$ pointing down in the vertical direction is used in this simulation. Lastly, a separate simulation on linear elastic materials is run to obtain the equilibrium initial stresses of each material point. Note that SI units are used in this study instead of the Imperial units used in previous studies.

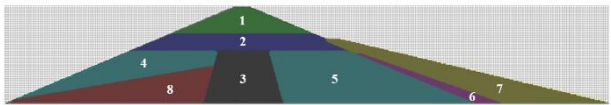

Figure 6. The different materials in the initial configuration of the MPM model.

Table 1. Summary of the strength parameters of the different materials.

| Number | Soil Type | $\phi$ | c (kPa) |
|---|---|---|---|
| 1 | Rolled Fill | 31 | 14.4 |
| 2 | Ground Shale | 27 | 24.0 |
| 3 | Clay Core | 0 | 30.0 |
| 4 | Upper Hydraulic Fill | 35 | 0.001 |
| 5 | Lower Hydraulic Fill | 35 | 0.001 |
| 6 | Drain | 37 | 0.001 |
| 7 | Berm | 37 | 4.8 |
| 8 | Liquefied Soil | 0 | 2.0 |

## 4.2 Results and Discussion

The evolution of the dam failure with time can be observed in Figure 7. The final crest loss in the MPM simulation is observed to be around 11 m, which is consistent with the observed event and the findings of Chowdhury (2019). The final runout stops at 43 m to the upstream direction, which is close to the approximate 42 m observed runout length of the section in the event. This is an advantage of the MPM since the FDM results in Figure 5 cannot achieve such large deformation, even with re-meshing.

In addition, there are a few more observations that can be made about the results. First, at 10 s, failure planes are visible within the clay core in both upstream and downstream directions. Nevertheless, the failure plane progresses more rapidly in the upstream direction as observed through the higher equivalent plastic deviatoric strain at 20 s. Second, the contour resembles some blocky feature that are observed in the event. Further research is needed to obtain more realistic blocky feature, which may include applying localization methods and softening material models within the MPM.

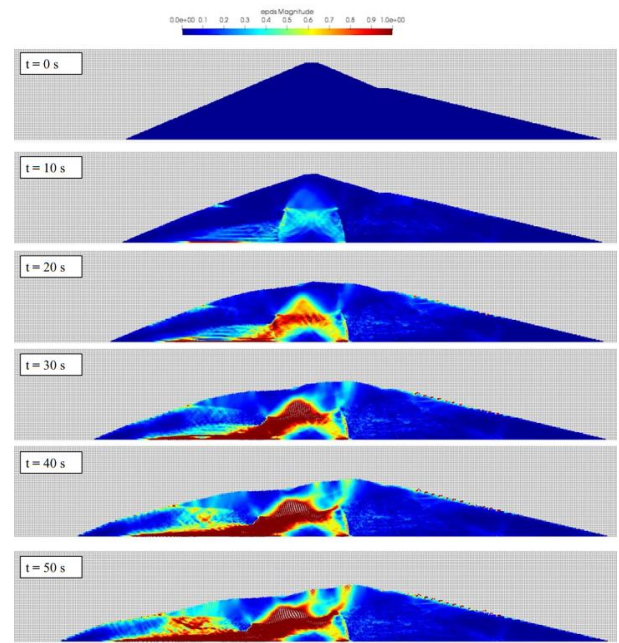

Figure 7. The evolution of the dam failure over time (upstream is to the left in this figure). The contour is showing the equivalent plastic deviatoric strain, where red indicates higher value.

## 5 REMARKS AND FUTURE STUDY

This study has shown the applicability of the MPM to simulate a specific large deformation problem, that is the liquefaction-induced dam failure of the Lower San Fernando Dam. This is an advantage of the MPM over the current state of the art methods including the FDM and the FEM.

There are a few recommended future works that can be built upon the findings of this study. The first is the overall simulation using MPM including the dynamic simulation using appropriate ground motion. The second is the use of more advanced constitutive model that can capture cyclic behavior of liquefied soil such that can be found in this case study. Lastly, further analysis using effective-stress based multiple-phase MPM that involves water and potentially, air, can prove to be useful to improve this simulation as well as gaining insights on the failure mechanism and runout behavior.